\documentclass[a4paper,11pt]{article}
\usepackage{aaskaiid}
 
\usepackage[normalem]{ulem}
\usepackage{xcolor}
\usepackage{physics}
\usepackage{amsmath}
\usepackage{bm}
\usepackage{comment}
\usepackage{ulem}
\usepackage{soul}
\usepackage{aaskaiid}
\usepackage{orcidlink}

\definecolor{darkred}{RGB}{175,0,0}

\def\vk{{\bm{k}}}

\def\hvk{{\hat{\bm{k}}}}
\def\ps{{\cal P}_{_{\rm S}}}
\def\As{A_{_{\rm S}}}
\def\fnl{f_{_{\rm NL}}}
\def\gnl{g_{_{\rm NL}}}
\def\tgnl{\tilde{g}_{_{\rm NL}}}

\def\ogw{\Omega_{_{\mathrm{GW}}}}

\newcommand{\beq}{\begin{equation}}
\newcommand{\eeq}{\end{equation}}

\newcommand{\ho}{\hat{\Omega}}

\newcommand{\pcite}[1]{~\cite{#1}}

\title{Exploring Gravitational Wave Signatures Due to Primordial Non-gaussianity and Large Scale Structure Using SKAO}
\ShortTitle{GW signatures due to primordial non-Gaussianity and LSS}

\author[1,2]{H.~V.~Ragavendra\orcidlink{0000-0001-7107-7345}}
\ShortName{Ragavendra et al.} 
\emailAdd{ragavendra.hv@pd.infn.it}
\author[3,4]{Giulia Capurri\orcidlink{0000-0003-0889-1015}}
\emailAdd{giulia.capurri@df.unipi.it}
\author[1,2]{Alina Mierna\orcidlink{0009-0009-0341-3525}}
\emailAdd{alina.mierna@studenti.unipd.it}
\author[1,2,5]{Gabriele Perna\orcidlink{0000-0002-7364-1904}}
\emailAdd{gabriele.perna@phd.unipd.it}
\author[1,2]{Federico Semenzato\orcidlink{0009-0006-0219-6192}}
\emailAdd{federico.semenzato.1@phd.unipd.it}
\author[1,2,6]{Nicola Bartolo\orcidlink{0000-0001-8584-6037}}
\emailAdd{nicola.bartolo@pd.infn.it}
\author[1,2,6]{Daniele Bertacca\orcidlink{0000-0002-2490-7139}}
\emailAdd{daniele.bertacca@unipd.it}
\author[3]{Lorenzo Valbusa Dall'Armi\orcidlink{0000-0002-0412-8058}}
\emailAdd{lorenzo.valbusadallarmi@df.unipi.it}
\author[1,2,6,7]{Sabino Matarrese\orcidlink{0000-0002-2573-1243}}
\emailAdd{sabino.matarrese@pd.infn.it}
\author[1,2,8]{Matteo Pegorin\orcidlink{0009-0003-1248-871X}}
\emailAdd{matteo.pegorin.1@phd.unipd.it}
\author[3,4]{Angelo Ricciardone\orcidlink{0000-0002-5688-455X}}
\emailAdd{angelo.ricciardone@unipi.it}

\affiliation[1]{Dipartimento di Fisica e Astronomia ``Galileo Galilei" 
Università di Padova, Via Marzolo 8, I-35131 Padova, Italy}
\affiliation[2]{Istituto Nazionale di Fisica Nucleare (INFN), Sezione di Padova, Via Marzolo 8, I-35131 Padova, Italy}
\affiliation[3]{Dipartimento di Fisica ``Enrico Fermi", Università di Pisa, Largo Bruno Pontecorvo 3, Pisa I-56127, Italy}
\affiliation[4]{Istituto Nazionale di Fisica Nucleare (INFN), Sezione di Pisa, Largo Bruno Pontecorvo 3, Pisa I-56127, Italy}
\affiliation[5]{National Institute for Chemical Physics and Biophysics (NICPB), Ravala 10, 10143 Tallinn, Estonia}
\affiliation[6]{Istituto Nazionale di Astrofisica (INAF), Osservatorio Astronomico di Padova, vicolo dell’ Osservatorio 5, I-35122 Padova, Italy}
\affiliation[7]{Gran Sasso Science Institute, I-67100 L’Aquila, Italy}
\affiliation[8]{Max Planck Institute for Gravitational Physics (Albert Einstein Institute), Am M{\"u}hlenberg 1, Potsdam 14476, Germany}

\abstract{This chapter explores theoretical and observational strategies to use the stochastic gravitational-wave background detectable by Square Kilometre Array Observatory (SKAO) as a probe of precision cosmology. We detail the critical phenomenon of scalar-induced gravitational waves, demonstrating their unique features and their sensitivity to primordial non-Gaussianity on scales much smaller than those probed by the cosmic microwave background and large-scale structure. 
We investigate the phenomenology of parity violation in the early Universe through the chirality imprinted in the stochastic gravitational-wave background, demonstrating that a parity-odd primordial trispectrum can generate a detectable scale-dependent helicity.
On the observational side, we point out that the standard gravitational-wave angular auto-correlation analysis is significantly limited by astrophysical shot noise. 
We show that cross-correlating the gravitational wave signal with independent large-scale structure tracers enhances the signal-to-noise ratio and allows us to isolate the astrophysical and cosmological components in the background.
These results can be achieved only thanks to the enhanced sensitivity of SKAO, extensive sky coverage, and high angular resolution, which together make such targets observationally feasible.}

\begin{document}
\maketitle

\newcommand{\actaa}{Acta Astron.} 
\newcommand{\araa}{ARA\&A} 
\newcommand{\aar}{A\&ARv} 
\newcommand{\aapr}{A\&ARv} 
\newcommand{\ab}{Astrobiol.} 
\newcommand{\aj}{AJ} 
\newcommand{\apj}{ApJ} 
\newcommand{\apjl}{ApJL} 
\newcommand{\apjs}{ApJSS} 
\newcommand{\ao}{Appl. Opt.} 
\newcommand{\apss}{Astro. \& Space Sci.} 
\newcommand{\aap}{A\&A} 
\newcommand{\aaps}{A\&AS.} 
\newcommand{\baas}{Bull. Am. Astron. Soc.} 
\newcommand{\caa}{Chinese A\&A} 
\newcommand{\cjaa}{Chinese J. A\&A} 
\newcommand{\cqg}{Class. Quantum Gravity} 
\newcommand{\gal}{Galaxies} 
\newcommand{\gca}{Geo. Cosmo. Acta} 
\newcommand{\icarus}{Icarus} 
\newcommand{\jcap}{JCAP} 
\newcommand{\jgr}{J. Geophys. Res.} 
\newcommand{\jgrp}{J. Geophys. Res. Planets} 
\newcommand{\jqsrt}{J. Quant. Spectrosc. Radiat. Transf.} 
\newcommand{\memsai}{Mem. SAIt} 
\newcommand{\mnras}{MNRAS} 
\newcommand{\nat}{Nature} 
\newcommand{\nastro}{Nat. Astron.} 
\newcommand{\ncomms}{Nat. Commun.} 
\newcommand{\nphys}{Nat. Phys.} 
\newcommand{\na}{New Astron.} 
\newcommand{\nar}{New Astron. Rev.} 
\newcommand{\physrep}{Phys. Rep.} 
\newcommand{\pra}{Phys. Rev. A} 
\newcommand{\prb}{Phys. Rev. B} 
\newcommand{\prc}{Phys. Rev. C} 
\newcommand{\prd}{Phys. Rev. D} 
\newcommand{\pre}{Phys. Rev. E} 
\newcommand{\prx}{Phys. Rev. X} 
\newcommand{\prl}{Phys. Rev. Let.} 
\newcommand{\psj}{Planet. Sci. J.} 
\newcommand{\planss}{Planet. Space Sci.} 
\newcommand{\pnas}{Proc. Natl Acad. Sci. USA} 
\newcommand{\procspie}{Proc. SPIE} 
\newcommand{\pasa}{PASA} 
\newcommand{\pasj}{PASJ} 
\newcommand{\pasp}{PASP} 
\newcommand{\rmxaa}{RMXAA} 
\newcommand{\sci}{Science} 
\newcommand{\sciadv}{Sci. Adv.} 
\newcommand{\solphys}{Sol. Phys.} 
\newcommand{\sovast}{Soviet Ast.} 
\newcommand{\ssr}{Space Sci. Rev.} 
\newcommand{\uni}{Universe} 

\setlength{\bibsep}{0.0pt} 

\tableofcontents

\section{Introduction}

The stochastic Gravitational Wave Background (GWB), arising from the isotropic (and anisotropic) superposition of unresolved gravitational wave sources, constitutes a fundamental probe of both primordial cosmology and the cosmic evolution of astrophysical systems (see~\cite{Sathyaprakash:2009xs,Bartolo:2016ami,Caprini:2018mtu,LISACosmologyWorkingGroup:2022jok} and references therein).
Current data from the Pulsar Timing Array (PTA) collaboration have provided the first existence of a stochastic signal in the nano-Hertz band~\cite{NANOGrav:2023gor,EPTA:2023fyk,NANOGrav:2023tcn,NANOGrav:2023hfp,NANOGrav:2023hvm}.
The transition from detection to precise characterization, however, requires a rigorous theoretical understanding of its complex signatures and the development of advanced observational strategies, particularly with the advent of next-generation probes such as SKAO~\cite{Janssen:2015,Keane:2014vja}.
This chapter explores the theoretical and observational frontiers of science enabled by the GWB, accessible to SKAO, in both the AA4 and Beyond-AA4 configurations.  
We focus on three distinct yet interrelated aspects: the phenomenon of Scalar-Induced Gravitational Waves (SIGWs) and the role of primordial non-Gaussianity~\cite{Perna:2024ehx,Mierna:2024pkh}; 
the imprint of chirality in SIGWs arising from parity violation mechanisms in the early Universe~\cite{Ragavendra:2025svk}; 
and the robust exploitation of stochastic GWB anisotropies through cross-correlation with large-scale structure (LSS)~\cite{semenzato2024GWBLSS,Sah_2024}.

The chapter is comprised of three sections detailing the three aspects
mentioned above, and concludes with a brief summary.
Section 2 discusses the phenomenon of SIGWs~(see \cite{Domenech:2021ztg} for a review, and references therein). 
Generated at second order from the coupling of primordial scalar and tensor perturbations, SIGWs are intrinsically linked to the spectral features of scalar perturbations~(see, for early works~\cite{Tomita:1967wkp,Matarrese:1992rp,Matarrese:1993zf}).
SIGWs could constitute a non-negligible fraction of the stochastic GWB in the PTA data (see for e.g.,~\cite{Powell:2019kid,Figueroa:2023zhu}).
We detail how the presence of primordial non-Gaussianity at the level of bispectrum, quantified through the non-linearity parameter $\fnl$~\cite{Planck:2019kim},
can fundamentally alter the spectrum of the SIGW signal $\ogw(f)$~\cite{Adshead:2021hnm,Ragavendra:2021qdu,Perna:2024ehx}.
These non-Gaussian contributions, while proportional to higher powers of the scalar amplitude, 
become significant in scenarios featuring enhanced scalar spectrum on small scales (for e.g., ref~\cite{Cai:2018dig,Unal:2018yaa,Atal:2021jyo,Ragavendra:2020sop}). 
Through this behavior, SIGWs  provide a unique probe of inflationary dynamics on small scales that are inaccessible to Cosmic Microwave Background (CMB) and LSS measurements (for instance, see~\cite{Firouzjahi:2023lzg,Cai:2023dls,Fei:2023iel,Li:2023xtl}).

Section 3 extends this discussion by examining the characteristic signature of parity-violation in the form of chirality in cosmological GWB, i.e., the degree of asymmetry between the spectral densities of their helical polarizations~\cite{Belgacem:2020nda,Sato-Polito:2021efu,Orlando:2020oko}.
We show that a parity-odd component of the primordial scalar trispectrum, parameterized by $\tgnl$, induces a sizable preferential handedness in the SIGW signal~\cite{Ragavendra:2025svk}.
The degree of chirality is found to be highly scale-dependent. Moreover, over a range of scales close to the peak amplitude, chirality provides a direct and clean window into the ratio of parity-odd to parity-even components of scalar non-Gaussianity. 
Crucially, while the monopole (all-sky-average) of this chiral V-mode may be difficult to measure, the anisotropies of the stochastic GWB are shown to be a sensitive to it (see, for instance~\cite{Belgacem:2020nda,Sato-Polito:2021efu,Cruz:2024esk}).
This effect can be leveraged to impose constraints on $\tgnl$, thereby
offering complementary constraint on the parameter over the scales probed by surveys of LSS~\cite{Philcox:2022hkh,Hou:2022wfj}.

Section 4  addresses the transition from the theoretical prediction of the GWB to its observational characterization, focusing on its anisotropic component. In the nano-Hertz frequency band, the GWB is primarily sourced by the cosmic population of inspiraling Supermassive Black Hole Binaries (SMBHBs)~\cite{Sesana:2008mz, andrew_smbhb}. These binaries reside in massive galaxies, which in turn trace the Universe's LSS. Consequently, the intrinsic spatial clustering of these GW sources imprints a characteristic anisotropy onto the GWB~\cite{Mingarelli:2013, PolTaylorRomano:2022}. This anisotropy is statistically described by its angular power spectrum, $C_\ell$, which quantifies the correlation strength of the GW signal as a function of the angular scale $\ell$.
A direct measurement of this cosmological structure via the GWB auto-correlation power spectrum ($C_\ell^{\rm GW, GW}$) is severely contaminated by a stochastic shot-noise component arising from the finite and discrete number of sources. This shot noise can overwhelm the underlying signal imprinted by the cosmic clustering of sources.
To isolate the cosmologically meaningful LSS signal,~\cite{semenzato2024GWBLSS}  advocates for the necessary implementation of cross-correlation techniques~\cite{Sah_2024, Cusin:2025xle}. This method involves correlating the GWB anisotropy map with independent tracers of the LSS, such as maps derived from large-scale galaxy surveys. The resulting cross-power spectrum, $C_\ell^{\rm gal, GW}$, is sensitive only to the spatial pattern, common to both the GWB and the galaxy distribution. This approach allows us to extract the imprint of LSS on the GWB clustering signal, making the GWB a novel and powerful probe of the matter distribution and opening a new channel to study the properties of the SMBHB population.

The successful execution of these analyses critically depends on the capabilities of the SKAO. 
Current limitations of PTA including highly non-uniform sky response, large sample variance, and poor constraints on high-$\ell$ modes~\cite{konstandin2024impactcosmicvarianceptas}, 
will be mitigated by the SKAO improved configuration~\cite{Xin_2021}. The projected increase in the number of high-precision millisecond pulsars ($N_{\text{psr}} \sim 700$ for SKA1~\cite{Xin_2021} and more optimistic
forecasts considering even~$10^3$ pulsars~\cite{Janssen:2015})
will dramatically expand the number of measurable angular modes, 
and ensure a near-isotropic sky coverage which will provide an enhanced sensitivity to the anisotropic component. This advancement allow us to access  high-resolution GWB maps and probe their spatial structure, establishing the GWB as a robust cosmological probe of the aforementioned phenomena of primordial and late-times.

\section{Primordial sources of induced gravitational waves}
Primordial gravitational waves provide a powerful probe of the early Universe. Generated during inflation and the subsequent radiation-dominated era, they are highly sensitive to the detailed dynamics of cosmic expansion and can therefore test a wide range of cosmological scenarios (see \cite{Guzzetti:2016mkm,Caprini:2018mtu} for a review and references therein). After their generation, primordial gravitational waves propagate almost freely throughout the Universe due to the weakness of the gravitational interaction, preserving information about their sources~\cite{Guzzetti:2016mkm,Caprini:2018mtu}. This underlines the importance of detecting primordial GWs and justifies the effort in building GW detectors spanning over a wide frequency range. Different cosmological GW sources can produce a detectable signal in the nano-Hertz frequency band (see~\cite{Sathyaprakash:2009xs,Bartolo:2016ami,Caprini:2018mtu,LISACosmologyWorkingGroup:2022jok}. Among such cosmological signals, the guaranteed ones are the SIGWs~~\cite{Tomita:1975kj, Matarrese:1993zf, Acquaviva:2002ud, Mollerach:2003nq, Carbone:2004iv, Ananda:2006af, Baumann:2007zm,Domenech:2021ztg}. As the name suggests, these are GWs generated by a coupling between scalar and tensor perturbations, which arises at second order in perturbation theory. Scalar (i.e., temperature) fluctuations have already been observed in the CMB. This observation implies the existence of SIGWs, which are directly linked to the spectral properties of the scalar perturbations and could, in principle, be used to probe them. Additionally, as we will argue in the next subsection, these signals are highly sensitive to possible deviations from the Gaussian statistics of the scalar perturbations, making their detection even more appealing. 
Another possible source of GWs is represented by cosmic strings  \cite{Vachaspati:1984gt, Sakellariadou:1990ne, Berger:1989tw, Damour:2000wa, Damour:2001bk,Damour:2004kw, Blanco-Pillado:2017oxo}, which are topological defects that may form, for instance, after a phase transition in the early-Universe \cite{Vilenkin:1982ks,Kamionkowski:1993fg, Caprini:2007xq, Huber:2008hg, Hindmarsh:2015qta}.

In this section, we will illustrate that we can probe and constrain primordial scalar non-Gaussianity
at the level bispectrum using the spectral density of GWB probed by SKAO.
We discuss the phenomenon of SIGW and the associated prediction on the shape of spectral density of
GWB to levy constraint on the non-Gaussianity parameter $\fnl$.
We also discuss how additional information can be extracted from the anisotropies in the GW signal of cosmological origin, which shall help distinguish different sources of the signal.

Although the detection of a Hellings–Downs correlation pattern in pulsar timing residuals is a distinctive signature of a GW signal, it does not reveal its origin. The measured correlation, in general, is the result of a superposition of both a cosmological and an astrophysical background. This highlights the need for more accurate modelling of the expected spectra and the key role of SKAO, which will deliver additional data with significantly improved precision.

\subsection{Effect of primordial non-Gaussianity on the spectral density}
\begin{figure}
\centering
\includegraphics[width=0.49\linewidth]{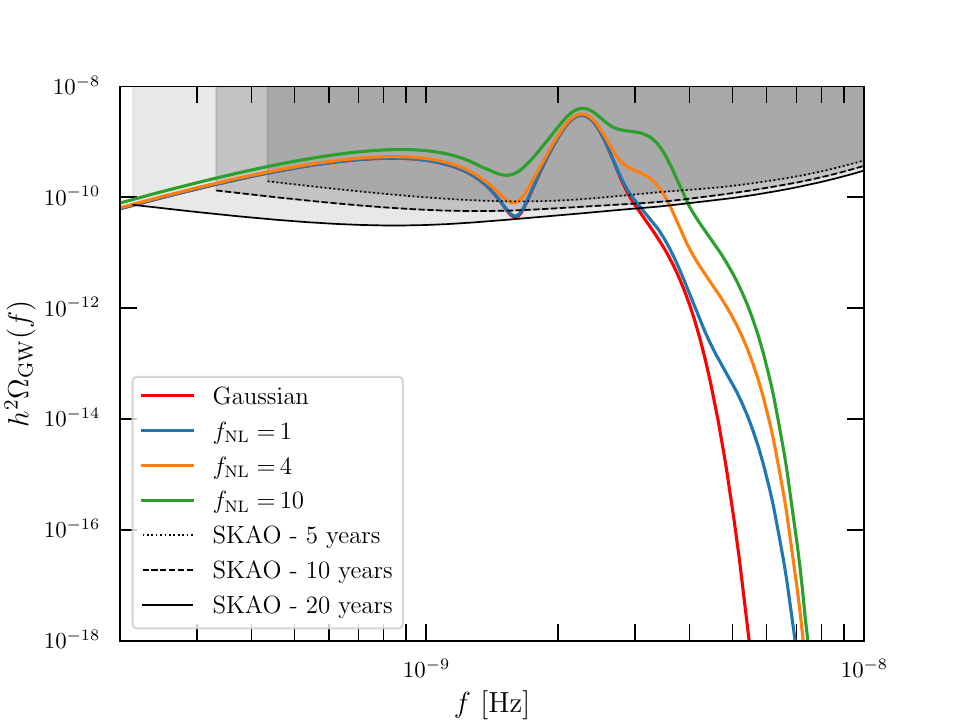}
\vskip -0.1in
\caption{Plot of the spectral energy density $h^2 \ogw$ for SIGWs in the Gaussian case (red solid line) as well as accounting for the presence of non-Gaussianity (blue, orange, green solid lines) in the case of a log-normal scalar power spectrum seed [adapted from~\cite{Perna:2024ehx}].}
\label{fig:ng-peak}
\end{figure}
Primordial non-Gaussianity quantifies the departure of the quantum fluctuations from the Gaussian behaviour, and it is considered a fundamental observable to shed light on the dynamics of the primordial Universe (e.g., see ~\cite{Bartolo:2004if,Babich:2004gb,Komatsu:2010hc}  for a review). Standard models of inflation, with a single field slow rolling along its potential~\cite{Guth:1982ec,Starobinsky:1982ee}, predict an amount of non-Gaussianity proportional to the slow-roll parameters, thus very small. However, different models can generate large non-Gaussian perturbations~\cite{Bartolo:2004if}. The Planck Collaboration has placed strong constraints for different types of non-Gaussianity, all of which are still compatible with zero ~\cite{Planck:2019kim}. Additionally, the error bars are large enough to allow for large values of non-Gaussianity. One could also argue that these constraints hold only on the largest scales, leaving the smallest ones unconstrained. In the case of a Gaussian field, its distribution is fully characterised by its first and second moments. In case of non-vanishing non-Gaussianity, instead, higher-order moments are crucial to characterize the distribution. The lowest order correlator is the three-point correlation function, which, in Fourier space, corresponds to the bispectrum. Given a generic field $\Phi$, it is defined as
\begin{equation}
    \langle \Phi_{\mathbf{k}_1}\Phi_{\mathbf{k}_2}\Phi_{\mathbf{k}_3}\rangle = (2\pi)^3\delta_{\rm D}(\mathbf{k}_1+\mathbf{k}_2+\mathbf{k}_3)B_{\Phi}(k_1,k_2,k_3)\,,
\end{equation}
where $B_\Phi$ is the bispectrum. Here, the Dirac delta imposes the triangle configurations for the wave-numbers due to the homogeneity of primordial perturbations.  The bispectrum can be written as \cite{Liguori:2010hx}
\begin{equation}
    B_\Phi(k_1,k_2,k_3) = f_{\rm NL} F(k_1,k_2,k_3)\,,
\end{equation}
where $f_{\rm NL}$ quantifies the amount of non-Gaussianity, while $F$ encodes the dependence on the specific triangular shape associated to the model considered. The main shapes considered by the Planck collaboration are the local, the equilateral and the orthogonal shape. Constraining them is thus necessary to shed light on the inflationary Universe~\cite{Bartolo:2004if,Planck:2018vyg}. 

Interestingly, the presence of primordial non-Gaussianity influences the production of GWs in both the early and late Universe in multiple ways. 
The SIGW spectrum is highly sensitive to presence of non-Gaussianity, being dependent by definition on the trispectrum of the primordial perturbations. In fact, one can relate the dimensionless SIGW energy density $\Omega_{\rm SIGW}$ to the trispectrum of scalar perturbations as $\Omega_{\rm SIGW}(k)\propto\int_{\mathbf{p}_1}\int_{\mathbf{p}_2}\langle\zeta_{\mathbf{p}_1}\zeta_{\mathbf{k}-\mathbf{p}_1}\zeta_{\mathbf{p}_2}\zeta_{\mathbf{k}-\mathbf{p}_2}\rangle$. As shown by~\cite{Adshead:2021hnm, Perna:2024ehx}, in the case of local non-Gaussianity, the corresponding GW spectrum shows additional contributions which can be schematically written at the emission as
\begin{equation}
    \Omega_{\rm SIGW}^{\rm TOT}(f) = \Omega_{\rm SIGW}^{\rm Gaussian} +     \Omega_{\rm SIGW}^{\rm non-Gaussian}\nonumber\\
    = A^2 \Omega_{\rm GW}^{\rm G}(f) + f_{\rm NL}^2 A^3 \Omega_{\rm GW}^{\rm NG}(f)
\end{equation}
where we indicated with $\Omega_{\rm GW}^{\rm G}(f)$ and $\Omega_{\rm GW}^{\rm NG}(f)$ the spectral shapes of the spectrum without any multiplicative factor and $A$ is the amplitude of the scalar power spectrum, which is constrained to be of order $\sim10^{-9}$~\cite{Planck:2018vyg} at CMB scales. In the case of scale-independent non-Gaussianity it is expected that the non-Gaussian contribution is small, being proportional to $A^3$. However on smaller scales, some mechanisms allow for an enhancement of the scalar power spectrum, which allows the amplitude to be of order $\sim 10^{-2}$,
thereby proportionately increasing the non-Gaussian contribution. 
As shown, for example, in~\cite{Figueroa:2023zhu}, PTA observations can be employed to constrain the value of $f_{\rm NL}$, with significant implications for models of inflation, suggesting that SKAO will have a great potential to improve the constraints. Even a non-observation of this signal would allow to place stringent upper bounds on the model parameters~\cite{Iovino:2024tyg}, possibly ruling out some models of inflation. 

In addition, the SIGW spectrum is model dependent, in the sense that its enhancement as well as its shape depend on the underlying assumptions on the inflationary dynamics. For what concerns the non-Gaussian part, the theoretical analysis is limited on considering local non-Gaussianity, since it allows to perform analytical and numerical calculations more easily. However, even different shapes and template for primordial non-Gaussianity could produce relevant imprints on the GW signal, which are yet to be explored.  This highlights that
more data and analysis are needed first for a better understanding of the cosmological signals but also to disentangle them  from the ones of astrophysical origin.
Besides, we may note that the sign of $\fnl$ remains unconstrained in this method as $\Omega_{\rm SIGW}$
is insensitive to it.

On the astrophysical side, primordial non-Gaussianity affects the clustering properties of the sources generating the GW. The main imprints are expected on the largest scales: in presence of local non-Gaussianity, in fact, the bias, which links the clustering of the GW sources to the underlying Dark Matter distribution, acquires a scale dependence whose amplitude is controlled by $f_{\rm NL}$ but also by the evolution history of the halo, as well as on its properties~\cite{Dalal:2007cu,Matarrese:2008nc,Slosar:2008hx}. One can write the non-Gaussian bias as
\begin{align}
    b_{\rm NG} = \bar{b} + C(k,z) \frac{b_\phi f_{\rm NL}}{k^2}
\end{align}
with $\bar{b}$ the Gaussian bias, and $C(k,z)$ a function (note that on large scales the dependence of this function on $k$ is negligible). From this relation it is clear how primordial non-Gaussianity enters the game but also how this effect is mainly relevant on the largest scales (small $k$), affecting thus only the lowest multipoles and being potentially detectable by SKAO, given its better angular resolution. A main limitation of this approach arises from our incomplete knowledge of the evolution of the underlying distribution, as well as from the uncertainty in how the linear bias itself is related to the formation and evolution of the halo hosting the binary~\cite{Lazeyras:2022koc,Barreira:2022sey}.

\subsection{Anisotropies of the primordial GWs}

Anisotropies of the cosmological GW  background could provide a crucial information about the source of GWs. The Boltzmann equation characterizes the evolution of GWs in the geometric optics limit and, similar to CMB, can be used to compute the linear anisotropy~\cite{Alba:2015cms, Contaldi:2016koz,Bartolo:2019oiq,Bartolo:2019yeu}.  We can define a graviton distribution function $f_{\rm GW}\left(\eta, \Vec{x}, \Vec{q}\right)$ as a function of the conformal time $\eta$, the position $\Vec{x}$ and the comoving momentum $\Vec{q}$ $(q= |\Vec{p}| a)$, which obeys the Boltzmann equation 
\begin{equation}
 \mathcal{L}\left[f_{\rm GW}\right] = \mathcal{C}\left[f_{\rm GW}\right] + \mathcal{I}\left[f_{\rm GW}\right] \, , 
\end{equation}
where $\mathcal{L}\left[f\right] = d f/d\lambda $ is the Liouville operator, $\mathcal{C}\left[f\right]$ and $\mathcal{I}\left[f\right]$ are the collision and emissivity terms  respectively. The collisional term
is absent for GWs and the emissivity term can be considered as an initial condition for the graviton distribution function. We consider a Friedmann-Lemaître-Robertson-Walker (FLRW) metric perturbed up to first order in the Poisson gauge 
\begin{equation}
ds^2 = a^2\left[-e^{2\Psi}d\eta^2 +e^{-2\Phi}\left(e^{\gamma}\right)_{ij}dx^idx^j\right],
\end{equation}
where  $\Phi$ and $\Psi$ are the large-scale scalar perturbations and $\gamma_{ij}$ denotes the transverse-traceless tensor perturbation. In the shortwave approximation, the tensor perturbations can be split into the small-scale perturbations that we identify as GWs $h_{ij}$, whose wavelength is much smaller than the scales on which the large-scale background tensor perturbations $H_{ij}$ vary and therefore $\gamma_{ij} = h_{ij} + H_{ij}$ \cite{Isaacson:1967sln}. The graviton distribution function $f_{\rm GW} $ can be expanded as the isotropic component $\bar{f}_{\rm GW} $ that solves the Boltzmann equation at zero order, $d\bar{f}_{\rm GW}/d\eta = 0$, and an anisotropic contribution that solves the Boltzmann equation at first order and can be parameterized as 
 \begin{equation}
    \delta f_{\rm GW} = -q\, \frac{d\bar{f}_{\rm GW}}{dq}\, \Gamma \, .
\end{equation}
where $\Gamma$ is a dimensionless function, which can be related to the perturbation in the GW energy density $\delta_{\rm GW} = \left[4-n_{\rm gwb}\right]\Gamma$ with $n_{\rm gwb}$ being the Gravitational Wave Background (GWB) spectral index. The solution of the Boltzmann equation written in terms of the function  $\Gamma$ in Fourier space is
\begin{equation}
    \begin{split}
        \Gamma(\eta_0,\vec{k},\hat{n},q) = &\Gamma(\eta_{\rm in},\vec{k},\hat{n},q)e^{ik\mu(\eta_{\rm in}-\eta_0)}+\Psi(\eta_{\rm in},\vec{k})e^{ik\mu(\eta_{\rm in}-\eta_0)}\\
        &+\int_{\eta_{\rm in}}^{\eta_0} d \tilde{\eta} \left[\Phi^\prime(\tilde{\eta},\vec{k})+\Psi^\prime(\tilde{\eta},\vec{k})-\frac{1}{2}\hat{n}^i \hat{n}^j H_{ij}^\prime(\tilde{\eta},\vec{k})\right]e^{ik\mu\tilde{\eta}} \, , 
    \end{split}
\end{equation}
where $\eta_{\rm in}$ is the initial time and $\mu\equiv \hat{k}\cdot \hat{n}$. The first term is due to the initial conditions and other two terms are due to the propagation of GWs through the large-scale scalar and tensor perturbations of the Universe. The initial condition depend on the specific source of GWs and can be related to the perturbations of other particle species or not if generated by independent degrees of freedom~\cite{Schulze:2023ich,
ValbusaDallArmi:2023nqn, ValbusaDallArmi:2024hwm, Mierna:2024pkh}. Given the better angular resolution of SKAO compared to current PTA observations, the initial anisotropy of the GW background can be an effective tool to distinguish among different sources in the nano-Hertz frequency range.

\section{Probing primordial parity violation via chirality in GWB}

Chirality in the cosmological GWB is a tell-tale signature of violation of parity in the universe. Several models of early universe have been explored in the literature to generate chiral GW, with the intensity of one helicity dominating over the other,
such as axion models of inflation (for e.g.~\cite{Barnaby:2010vf,Bartolo:2014hwa,Thorne:2017jft}), 
presence of parity-violating spectator fields (for e.g.~\cite{Adshead:2013qp,Anber:2016yqr}), 
primordial magnetogenesis (for instance~\cite{Caprini:2014mja,Kahniashvili:2020jgm,Okano:2020uyr,Brandenburg:2021bfx,RoperPol:2021xnd}) and 
certain modified gravity models (for e.g.~\cite{Takahashi:2009wc,Cai:2021uup,Bari:2023rcw,Feng:2024yic,Alexander:2024klf}).
Such helical imbalance, if found in GWB contributed by scalar-induced tensor perturbations as discussed in the preceding section, can be a probe of parity-violation in the higher-order correlations of the primordial scalar perturbations.
As shown in~\cite{Ragavendra:2025svk}, a unique source of chirality in SIGW is the  parity-odd component of the primordial scalar trispectrum. 
The amount of chirality thus induced is directly proportional to the non-Gaussianity parameter determining the strength of the parity-odd trispectrum $\tgnl$.

In this section, we illustrate a method to probe parity-violation in the primordial Universe using SKAO. Since the chirality of SIGW is sourced by the parity-odd part of the primordial trispectrum, the V-mode of GWB constrains the strength of the parity-odd trispectrum $\tgnl$. 
While the timing residuals of PTA may be insensitive to the monopole of the V-mode, they are sensitive
to the anisotropies and so shall serve as observables to constrain primordial parity-violation.

We may illustrate the effect of chirality being induced by the scalar trispectrum, using the following simple template of trispectrum~\cite{Shiraishi:2016mok,Shiraishi:2016hjd,Philcox:2025bvj}
\begin{subequations}
\begin{align}
{\cal T}_{\rm odd}(\vk_1,\vk_2,\vk_3,\vk_4) & = i\,\tgnl\, 
\beta(\widehat{\vk_1+\vk_2}, \hvk_1, \hvk_3)
P(k_1)P(k_3)P(\vert\vk_1+\vk_2\vert) 
+~\text{23 permutations}\,, \\
{\cal T}_{\rm even}(\vk_1,\vk_2,\vk_3,\vk_4) & = 2\,\gnl\, 
P(k_1)P(k_3)P(\vert\vk_1+\vk_2\vert) +~\text{11 permutations}\,,
\end{align}
\label{eq:trispec-paramet-1}
\end{subequations}
where $\gnl$ and $\tgnl$ are the strengths of even and odd components of the trispectrum.
The quantity $P(k) = 2\pi^2\,\ps(k)/k^3$, is the power spectrum of primordial scalar perturbations.
Since $\Omega_{\rm SIGW} \propto \langle\zeta_{\vk_1}\zeta_{\vk_2}\zeta_{\vk_3}\zeta_{\vk_4}\rangle$, the effect of parity-odd trispectrum is directly captured in the spectral density of $V$ mode, i.e. the Stokes parameter quantifying the degree of circular polarization.
As mentioned in the preceding section, there is a level of model dependence in this analysis. We work with a template of exchange trispectrum that is constituted by local-type three-point interaction~\cite{Philcox:2025wts}. But the choice illustrates the effect without loss of generality.

The degree of chirality in GWB is given by~\cite{Satoh:2010ep,Bartolo:2020gsh}
\begin{eqnarray}
\Pi_k &=& \frac{\ogw^V(k)}{\ogw(k)}
= \frac{\ogw^L(k) - \ogw^R(k)}{\ogw^L(k) + \ogw^R(k)}\,,
\end{eqnarray}
where $\ogw$ is the dimensionless spectral density of GW, 
$\ogw^V$ is the corresponding quantity for the $V$ mode of GW, 
$\ogw^{L(R)}$ is the spectrum associated with the left (right) circular polarization mode.
For a simple case of a scale-invariant, primordial scalar power spectrum $\ps(k) = \As$ substituted in the above expression of $\cal T$, we may estimate $\Pi_k$ to be
\begin{eqnarray}
\Pi_k &\simeq& \frac{7}{2\pi} \tgnl\,\As\,.
\end{eqnarray}

For the case of a peak in the scalar power, which is relevant for enhanced amplitudes of SIGW over small scales, we shall have
\begin{align}
\ps(k) &= \As\left(\frac{k}{k_\ast}\right)^{n_{\rm s} -1}
+ \frac{A_{\rm p}}{\sqrt{2\pi\sigma^2_{\rm p}}}
\exp\left[-\frac{1}{2\sigma_{\rm p}^2}
\ln^2\left( \frac{k}{k_{\rm p}} \right)\right]\,,
\label{eq:ps-peak}
\end{align}
where $\As$ and $n_{\rm s}$ ensure the right amplitude and shape of spectrum over
large scales of CMB around the pivot scale of $k_\ast = 0.05\,{\rm Mpc}^{-1}$.
The term with the lognormal form leads to an enhanced amplitude of 
$A_{\rm p}$ at $k_{\rm p} \gg k_\ast$.
\begin{figure}
\centering
\includegraphics[width=0.49\linewidth]{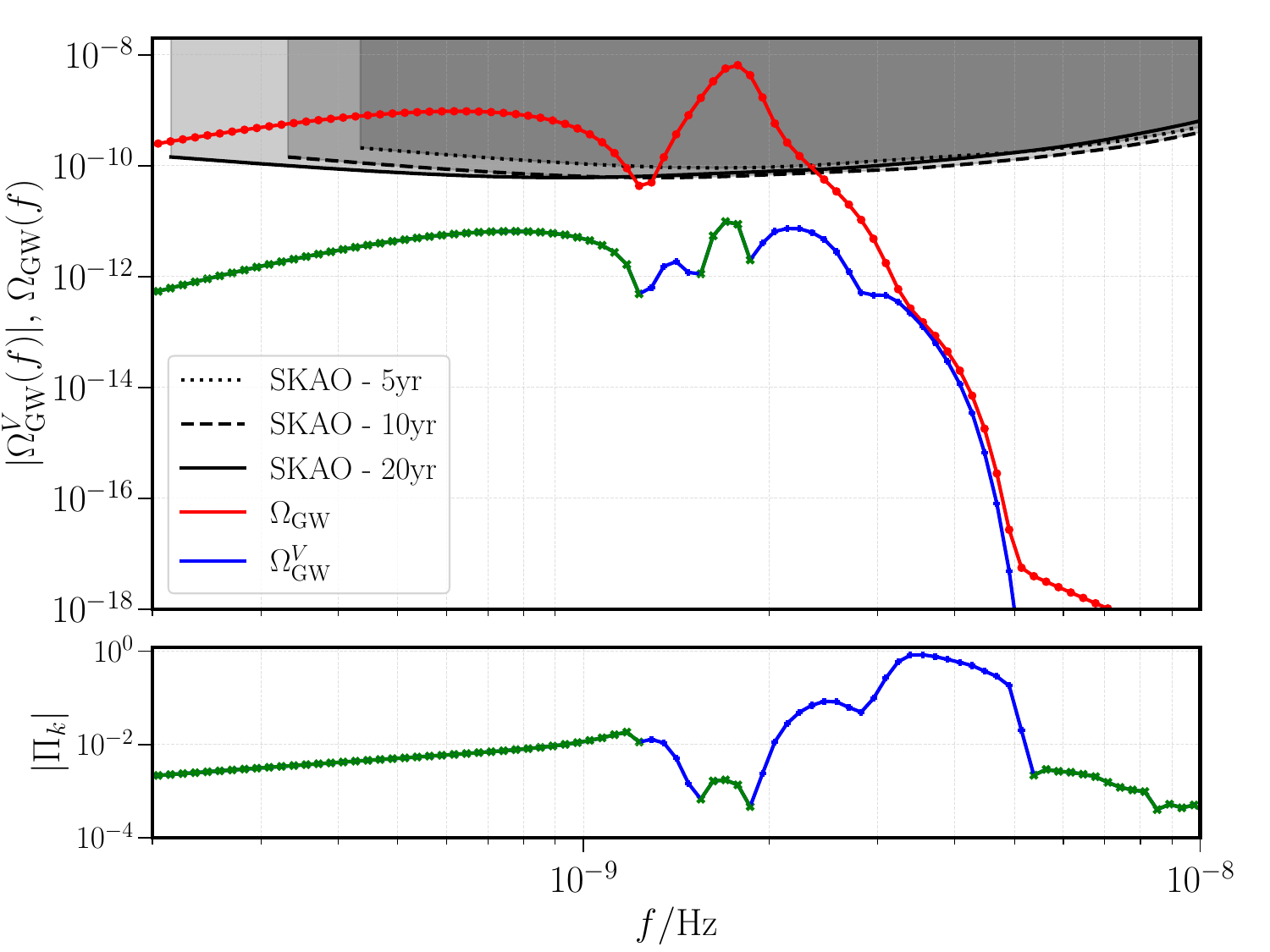}
\vskip -0.1in
\caption{We illustrate the spectral densities $\ogw$ and $\ogw^V$ of SIGW, 
while accounting for $\cal T$ given by Eq.~\eqref{eq:trispec-paramet-1}.
The scalar power spectrum $\ps(k)$ in this case is described by a lognormal peak [Eq.~\eqref{eq:ps-peak}].
We present $\ogw$ and $\ogw^V$ in red and blue curves respectively, and
indicate negatives values of $\ogw^V$ in green.
We present the corresponding degree of chirality $\Pi_k$ in the panel below. 
We have set $A_{\rm p}=10^{-2}$, $\gnl=\tgnl=1$ and $\sigma^2_{\rm p} = 10^{-2}$.
Setting $k_{\rm p} = 10^6\,{\rm Mpc}^{-1}$ (corresponding to 
$f_{\rm p} \simeq 1.55\times 10^{-9}\,{\rm Hz}$), we obtain the signal in the 
range of frequencies observable by SKAO.
We have also included the sensitivity of SKAO to $\ogw$ through a PTA style observation over different durations, in solid, dashed and dotted black curves~\cite{Shannon01.2026.SKA}.
We observe the following behaviors from this analysis:
$\Pi_k$ is highly scale-dependent, changing signature around the peak 
($k \simeq k_{\rm p}$).
Notably, $\Pi_k$ is significantly large over the wavenumbers of 
$k \simeq 3k_{\rm p}$, reaching as high as $10\%-100\%$~
[adapted from~\cite{Ragavendra:2025svk}].}
\label{fig:ph-peak}
\end{figure}
We present the behavior of $\ogw^V(k)$ and $\ogw(k)$ in this case along with their ratio $\Pi_k$ in Fig.~\ref{fig:ph-peak}.
We have chosen $k_{\rm p} = 10^6\,{\rm Mpc}^{-1}$, which corresponds to 
$f_{\rm p} \simeq 1.55\times 10^{-9}\,{\rm Hz}$, the frequency range of interest for SKAO~(see~\cite{Ragavendra:2021qdu} for inflationary models generating peaks in this frequency range).
The overlap of sensitivity curves of SKAO to $\ogw$ for different observing times~\cite{Shannon01.2026.SKA} and the peak amplitude of the theoretically predicted $\ogw$ suggests the possibility of obtaining reliable parameter constraints for the scenario of interest.
We find that the corresponding chirality alters sign around the peak, i.e. $\Pi_k$ is negative (right-circular) 
away from the peak over $k \ll k_{\rm p}$ and positive (left-circular) just 
after the peak $k \gtrsim k_{\rm p}$.
Secondly, $\tgnl=1$ easily induces chirality of $\Pi_k \gtrsim 10^{-3}$
around the peak. Importantly, at $k \simeq 3k_{\rm p}$, $\Pi_k$ reaches as high 
as $\Pi_k \simeq 0.1-1$.

Focussing specifically over the region adjacent to $k_{\rm p}$, we find that the behavior of $\ogw$ around $k \simeq 3k_{\rm p}$ is dominated by contribution from ${\cal T}_{\rm even}$.
Here, the degree of chirality becomes
\begin{equation}
\Pi_{k \simeq 3k_{\rm p}} \simeq \frac{\tgnl}{\gnl}\,,
\end{equation}
the ratio of amplitudes of the parity-odd to parity-even parts of the trispectrum.
Thus, the degree of chirality of SIGW, over a range of scales adjacent to the peak, 
is a direct window to the degree of parity violation in the primordial scalar trispectrum.

Even in case of not detecting strong chirality, an upper bound on $\Pi_k$ will translate to a bound on $\tgnl$.
In case of the observed GWB being a composite signal of which SIGW is a constituent fraction along with other sources, possibly astrophysical in origin, the bound shall be relatively weaker proportional to the relative intensity of SIGW.
Nevertheless, we must emphasize that it shall be a unique probe of $\tgnl$ over small scales of around $10^6\,{\rm Mpc}^{-1}$, compared to probes of 
$\tgnl$ such as galaxy surveys over large scales of around $10^{-2}\,{\rm Mpc}^{-1}$.

A caveat of this analysis that we already noted is that, the monopole of the V-mode of GWB may not be measurable using the correlations of the time-delay residuals in the signals from an array of pulsars.
However, the anisotropies of the signal shall be sensitive to it~\cite{Belgacem:2020nda,Sato-Polito:2021efu,Cruz:2024esk,Chen:2024ikn,Caporali:2025dyf}. 
Thus, the measurement of GW with better resolution of anisotropies using a larger and widely-distributed
array of pulsars, shall enable constraining the chirality of the GW and in turn introduce a new bound 
on $\tgnl$.
Such a bound from SKAO shall complement the bounds on $\tgnl$ over large scales as is currently 
being inferred from CMB~\cite{Philcox:2025wts} and galaxy surveys~\cite{Philcox:2022hkh,Hou:2022wfj}, 
eventually informing us of the scale dependence of the trispectrum.


\section{Exploring cross-correlation of large-scale structure tracers with GWB}
The GWB in the nano-Hertz regime is expected to originate primarily from the cosmic population of inspiraling SMBHBs, which trace the formation and evolution of massive galaxies over cosmic time~\cite{Sesana:2008mz,Sesana:2010qb,M17,andrew_smbhb}. Since these binaries reside in galaxies, their distribution reflects the underlying large-scale structure (LSS) of the Universe, and the resulting GWB encodes information about it. 
In this Section, we review the potential of the SKAO to turn anisotropies of the nano-Hertz GWB from a largely theoretical target into a practically exploitable observable. 
In particular, SKAO will monitor a much larger number of pulsars compared to current PTAs, with improved timing precision, and better prospects for identifying and subtracting the brightest continuous-wave sources.
This unlocks the possibility of reconstructing  the angular structure of the nano-Hertz GWB sky, providing a new window to test the properties and ultimately the origin of the signal.
At the same time, the scientific return of this program will depend on how well SMBHB population uncertainties,  galaxy bias, resolved source subtraction, and redshift systematics can be controlled.

Several complementary techniques to characterize and measure these anisotropies with PTAs are being developed and validated~\cite{Mingarelli:2013,TaylorGair:2013,Gair:2014,AliHaimoud:2021,PolTaylorRomano:2022,Depta:2025}.
Cross-correlation techniques provide a powerful tool to extract this information by matching the GWB anisotropies with those of independent LSS tracers, such as galaxy counts or weak lensing maps~\cite{Sah_2024, semenzato2024GWBLSS, Cusin:2025xle}. This strategy can boost the detectability of the GWB, disentangle its astrophysical and cosmological components, and establish it as a novel tracer of the LSS.
Such cross-correlations are especially valuable as they are expected to be more robust than the GWB auto-correlation to shot noise from the discrete SMBHB population, and can therefore provide a cleaner test of whether the GWB follows the large-scale distribution of massive galaxies.
This approach has been extensively explored at higher frequencies, e.g., for stellar-origin binaries in the LIGO-Virgo-KAGRA or LISA bands. Recent studies show that, despite the presence of shot noise, such correlations remain detectable with next-generation instruments~\cite{Ricciardone:2021kel,Capurri:2021prz,ValbusaDallArmi:2022htu,Alonso:2020mva,Yang:2020usq,Canas-Herrera:2019npr, Scelfo_2018, Libanore:2020fim, LISACosmologyWorkingGroup:2022kbp, Bosi:2023amu}. Indeed, frameworks established at higher frequencies can be employed in the PTA regime, with adaptations related to, e.g., array geometry, noise properties, and likelihood modeling. 
However, a crucial difference arises in the nature of the observable. Each pair of pulsars is intrinsically sensitive to GWB power across all angular scales.
Therefore, this effect introduces its own potential bias, and can limit the maximum angular scale that can be probed ($\ell_{\rm max}$).
This should be taken into account in the process of reconstructing an unbiased of the GWB sky.

Forecasts indicate that future datasets, particularly from the SKAO, will enable statistically significant measurements of GWB–galaxy cross-correlations. With enhanced angular resolution and a growing pulsar network~\cite{Smits:2008,Janssen:2015,Xin_2021}, SKAO-era PTAs will allow high-confidence reconstructions of the GWB sky~\cite{Weltman:2020,PolTaylorRomano:2022,Depta:2025}, enabling cross-correlations with galaxy surveys such as LSST, Euclid, and DESI~\cite{Sah_2024, semenzato2024GWBLSS,Cusin:2025xle}. This will turn the anisotropic GWB in the nano-Hertz regime into a new cosmological observable, offering a GW counterpart to traditional electromagnetic probes of the LSS.

In this section, we briefly review the origin of GWB anisotropies in the PTA band, the advantages of cross-correlation techniques, and the current status and future prospects with SKAO.

\subsection{Spatial anisotropies of the Late-Universe GWB}

Multiple mechanisms can source GWB anisotropies, each carrying distinct astrophysical or cosmological signatures. 
While current PTA analyses often assume isotropy, a degree of anisotropy is expected from the superposition of contributing sources and their inherently non-uniform spatial distribution.
The detection and characterization of these anisotropies serves as a powerful discriminant between competing models of the background's origin. 
The dominant contribution arises form the incoherent superposition of signals from the population of inspiraling 
SMBHBs~\cite{Sesana:2008mz,M17,agazie2024_ACC,satopolito2024distributiongravitationalwavebackgroundsupermassive}. 
These SMBHBs reside in massive galaxies that themselves trace the LSS of the Universe. 
The GWB hence inherits this spatial clustering, with the angular distribution of GW power $P(\ho)$ encoding the same statistical features of the dark matter distribution mapped by galaxies. 
This fundamental connection makes GWB anisotropies a novel probe of the matter distribution, offering unique insights complementary to electromagnetic observations.

The connection between the astrophysical source properties and the PTA observable is formalized through the statistics of the timing residuals, $r_p(t)$, induced in each pulsar $p$. The fundamental quantity describing the background is the GW power spectrum dependence on sky direction, $H(f, \ho)$, where $f$ is the GW frequency and $\ho$ is the direction of propagation. For a stationary and unpolarized GWB, the two-point correlator of the GW strain Fourier amplitudes $h_A(f, \ho)$ is~\cite{Mingarelli:2013}:
\begin{equation}
    \langle h_A^*(f, \ho) h_{A'}(f', \ho') \rangle = \frac{1}{4} \delta(f-f') \delta_{AA'} \delta^{(2)}(\ho, \ho') H(f)P(\ho).
    \label{eq:h_correlator}
\end{equation}
Here, $H(f)$ represents the frequency-dependent part of the all-sky power, and $P(\ho)$ is the dimensionless angular power distribution, normalized such that it integrates to $4\pi$. 
The characteristic strain spectrum, related to the power spectral density by $h_c^2(f) = 4f H(f)$, follows the expected power law $h_c(f) = A_{\rm yr}(f/f_{\rm yr})^{-2/3}$ for a population of circular, GW-driven binaries, where the measured amplitude is $A_{\rm yr} = 2.4^{+0.7}_{-0.6} \times 10^{-15}$ at $f_{\rm yr} = 1~{\rm yr}^{-1}$~\cite{NANOGrav:2023gor}.
This sky distribution of power is not observed directly but is measured through its integrated effect on the cross-correlation of timing residuals between pairs of pulsars. The cross-power spectral density of the timing residuals, $\mathcal{R}_{pq}(f)$, for a pair of pulsars $p$ and $q$ is given by:
\begin{equation}
\mathcal{R}_{pq}(f) = \frac{H(f)}{(2 \pi f)^2} \int_{S^2} d\ho \, P(\ho) \, \Gamma_{pq}(\ho)
\end{equation}
where $\Gamma_{pq}(\ho)$ is the antenna beam pattern, or Overlap Reduction Function (ORF), for the pulsar pair. In the simplest case, neglecting the pulsar term (which is valid for $f D_p \gg 1$, where $D_p$ is the pulsar distance), the ORF encodes the geometric response of the detector to a GW from direction $\ho$. For an isotropic background where $P(\ho) = 1$, the integral encodes the Hellings and Downs curve\pcite{Mingarelli:2013}. 
For an anisotropic background, the measured correlation deviates from this curve in a way that depends on the multipole moments of $P(\ho)$\pcite{variance_HD,Grimm_2025}.
To reconstruct the GWB power angular distribution, $P(\ho)$ is typically decomposed into a basis of spherical harmonics, $P(\ho) = \sum_{\ell, m} c_{\ell m} Y_{\ell m}(\ho)$.
Further basis choices can be explored to optimize the reconstruction for specifc signal and detector configurations, see, e.g.,\pcite{Hotinli:2019tpc,alihaimoud2020,Taylor:2020zpk,Pol:2022sjn,NG15anisotropy}. 
The goal of an anisotropic search is to measure the coefficients $c_{\ell m}$ and ultimately  estimate the angular power spectrum $\hat{C}_\ell = 1/(2\ell+1)\sum_m |c_{\ell m}|^2$. However, the number of statistically independent modes that can be constrained is fundamentally limited by the number of distinct pulsar pairs in the array, $N_{\text{pairs}} = N_{\text{psr}}(N_{\text{psr}}+1)/2$.

\subsection{Probing LSS and SMBHB evolution with GWB cross-correlations}

The angular power spectrum of the total sky power $P(\ho)$ is a composite quantity that aggregates contributions from physically distinct phenomena. The primary component of cosmological interest is the correlated signal arising from the anisotropic clustering of SMBHB host galaxies tracing the LSS. 
An additional stochastic shot-noise component arises from the finite number and discrete nature of the unresolved sources contributing to the GWB. 
Moreover, a small population of individual, luminous continuous wave (CW) sources introduces an additional contribution \pcite{NNAOGRAV_CW,Gardiner2024,agazie2024_ACC,Agarwal:2025cag}. These sources are low-redshift massive binaries whose coherent emission stands out from the stochastic ensemble and can, in principle, be resolved individually.
Although all GWB anisotropies can be traced back to the LSS, these components possess distinct angular signatures and present a hierarchical challenge for data analysis. A single, bright CW source, as well as the collective shot noise, both contribute a nearly constant, scale-invariant power across all angular multipoles $\ell$\pcite{Agarwal:2025cag}. In contrast, the LSS-imprinted signal is expected to exhibit a non-trivial, scale-dependent structure reflective of the matter power spectrum and SMBHB distribution. 
The deterministic CWs and the stochastic shot noise act as foregrounds that can dominate and obscure the LSS signal. 
A robust measurement of the LSS imprint therefore requires a multi-step approach: first, the identification and subtraction of individually resolvable CW sources, and second, the careful modeling of the residual shot-noise level to isolate the underlying correlated signal.

The presence of these few luminous sources fundamentally alters the statistical nature of the GWB map. They introduce a high degree of variance between realizations and cause the statistics of the angular power spectrum coefficients, $\hat{C}_\ell$, to deviate significantly from the Wishart distribution expected for a Gaussian random field\pcite{semenzato2024GWBLSS}. This implies that the two-point function alone becomes an insufficient descriptor of the field, a critical consideration for current PTA experiments that have yet to resolve individual sources. By progressively lowering the resolvability threshold, $h_{\rm res}$, we effectively transition from a sky dominated by a few point-like sources to a true background sourced by a vast, unresolved population whose anisotropies are more representative of the LSS.
However, even after the removal of the brightest sources, a significant challenge remains. The auto-correlation of the residual GWB ($C_\ell^{\rm GW,GW}$) is still heavily contaminated by the shot noise from the discrete nature of the remaining population. The analysis in\pcite{semenzato2024GWBLSS} shows that this contamination is severe enough to render the auto-correlation statistically incapable of distinguishing a GWB that traces the LSS from one generated by a spatially uniform source distribution. 

The limitations of auto-correlation analysis can be overcome through the use of cross-correlation. 
The fundamental strength of this technique lies in its capacity to isolate a shared physical signal by leveraging two distinct and independent observational probes of the same underlying field\pcite{Dai:2015wla,Kovetz_2017,Scelfo_2018,Scelfo_2020}. In this context, both the galaxy distribution and the SMBHB-induced GWB are tracers of the same LSS. 
The cross-correlation angular power spectrum $C_\ell^{\rm gal,GW}$ can robustly distinguish the LSS-imprinted GWB from a uniform model at $1\sigma$, $2\sigma$, $3\sigma$, and $5\sigma$ significance with access to angular scales of $\ell_{\max} \gtrsim 20, 30, 42$, and $72$, respectively\pcite{semenzato2024GWBLSS}.  Achieving this requires a source-resolving sensitivity of $h_{\rm res} \approx 10^{-16}$, a capability anticipated for next-generation facilities like the SKAO. 

Beyond confirming the astrophysical origin of the GWB, cross-correlation with LSS tracers provides a unique probe of the cosmic evolution of the source-galaxy connection. While the auto-correlation signal is dominated by shot noise , the cross-spectrum carries a clean imprint of the SMBHB population's redshift distribution and its relation to host galaxy properties, such as the evolution of the $M_*-M_{\text{BH}}$ relation\pcite{Sah_2024}. 
Furthermore, in a controlled analytical model for the SMBHB distribution, forecasts show that GWB cross-correlations improve constraints on the redshift-dependent parameters governing SMBHB populations\pcite{Sah_2024,sah2025routeunveilcosmicgenealogy}.
This enables constraints on SMBHB cosmic merger history that are inaccessible through spectral measurements studies alone. 
Indeed, a measurement of the cross-correlation power spectrum can be used to constrain the redshift evolution, luminosity function, and host environment of SMBHBs. Moreover, it allows measurement of the gravitational-wave bias, i.e., the ratio between the clustering of the stochastic GWB and that of the matter field, which is crucial for understanding the relationship between GW sources and the underlying matter distribution.

However, precise modeling of the SMBHB population presents challenges. Uncertainties in stellar hardening, gas-driven migration, and eccentricity evolution can modify the residence time of binaries in the PTA band and hence the frequency spectrum and redshift distribution of the GWB \cite{Sesana:2010qb,Taylor_2016,Agazie_ecc}. Different prescriptions for SMBH growth and AGN/stellar feedback can alter the black-hole mass function and the SMBH-host connection, thereby changing the predicted SMBHB population and its cross-correlation with galaxies \cite{Sah_2024,sah2025routeunveilcosmicgenealogy}. A further source of uncertainty comes from the modeling of the large-scale-structure tracer itself: the interpretation of $C_\ell^{\rm gal,GW}$ depends on the tracer redshift distribution and bias, and can be impacted by photometric-redshift errors, magnification, and selection effects \cite{konstandin2024impactcosmicvarianceptas,Bertacca:2019fnt,Raccanelli:2015vla,Bernal:2020pwq}. For this reason, the most robust early SKAO-era applications will likely emphasize well-characterized tracers and angular scales where these systematics are best controlled.

Furthermore, the statistical properties of the GWB offer a powerful, although degenerate, window into the intrinsic dynamics of the binaries themselves. Specifically, the observed flattening of the GWB spectrum at low frequencies  can be attributed to either strong environmental coupling (e.g., efficient stellar hardening) or high orbital eccentricity of the binaries\pcite{Taylor_2016,Agazie_ecc}. These two scenarios can be disentangled by examining the GWB's frequency-correlated anisotropy\pcite{Truant_2025,moreschi2025dissectingnanohzgravitationalwave,fastidio2025realisticconsecutivegalaxymergers}. Eccentric binaries emit GWs over a broad spectrum of harmonics, meaning a single bright source will imprint a similar anisotropic pattern on the sky across multiple frequency bins. On the other hand, quasi-circular binaries are essentially monochromatic, leading to uncorrelated sky maps at different frequencies. High-resolution N-body simulations of realistic galaxy mergers suggest that SMBHBs are indeed expected to enter the PTA band with substantial eccentricities ($e>0.85$)\pcite{fastidio2025realisticconsecutivegalaxymergers}, making the search for such frequency correlations a physically motivated necessity. This internal cross-correlation of the GWB signal with itself therefore provides a key diagnostic to separate environmental effects from intrinsic binary dynamics, revealing a deeper layer of physical information about the final stages of SMBHB evolution.

\subsection{Current challenges and future prospects with SKAO}

The reconstruction of the GWB power distribution, $P(\ho)$, from timing residual data presents formidable observational and analytical challenges.

Current PTAs consist of a relatively small number of pulsars distributed anisotropically across the sky, leading to a highly non-uniform sky response where sensitivity is direction-dependent and certain angular modes are poorly constrained. Beyond these instrumental limitations, the coherent superposition of signals from all directions introduces interference terms that do not vanish in a single realization, leading to a sample variance in the measured cross-correlations that is significantly larger than the expectation from Gaussian statistics\pcite{konstandin2024impactcosmicvarianceptas}. This sample variance, compounded by the pulsar term and polarization interference, degrades the precision of any GWB map and can obscure the subtle, underlying physical signal we seek to measure.

The advent of the SKAO is poised to mitigate these challenges, transforming the landscape of GWB anisotropy science\pcite{Janssen:2015}. The primary advance will be a dramatic increase in the number of high-precision millisecond pulsars; conservative forecasts project $\sim$700-800 pulsars for SKA1 and SKA2, with optimistic estimates reaching over $10^3$ \pcite{Keane:2014vja,Xin_2021}. This increase in $N_{\text{psr}}$ directly expands the number of measurable angular modes, as the number of independent constraints scales with the number of pulsar pairs, $N_{\text{pairs}}$. Crucially, the SKAO will provide near-isotropic sky coverage, moving beyond the hemispheric biases of current arrays and enabling a more uniform and robust reconstruction of the GWB map. Furthermore, the large pool of available pulsars will permit the selection of an optimal subset with the lowest intrinsic red noise and highest timing precision, which will significantly enhance the signal-to-noise ratio for the faint anisotropic component. 
The MeerKAT Pulsar Timing Array already shows evidence for GWB anisotropies with 83 timed pulsars after just 4.5 years of data\pcite{Miles_2024}.
The combination of a larger, more isotropically distributed, and higher-fidelity pulsar array, is what makes the high-resolution maps required for the cross-correlation analyses a feasible goal for the next decade.

\section{Conclusions}

This chapter summarizes the science cases involving theoretical features and observational requirements that are necessary to promote the stochastic GWB from a promising detection to a high-precision cosmological and astrophysical probe. 

We have illustrated how the phenomenon of SIGW allows us to probe primordial scalar non-Gaussianity parameters namely $f_{\rm NL}$, $g_{\rm NL}$ and $\tilde g_{\rm NL}$ by studying the shape of $\Omega_{\rm GW}(f)$ and $\Omega^V_{\rm GW}(f)$ observable through SKAO~\cite{Perna:2024ehx,Ragavendra:2025svk}.
The constraints on these parameters over scales of $k=10^6\,{\rm Mpc}^{-1}$ corresponding to nano-Hertz frequency shall complement the constraints on them over large scales from CMB and galaxy surveys.
We also discussed the scale dependent bias of astrophysical GW sources and the imprints of primordial anisotropies in GWB that help distinguish between various contributions to the total GW signal probed by SKAO~\cite{Perna:2023dgg,Mierna:2024pkh}.
As to caveats of the outlined analyses, while the non-Gaussianity parameters can be easily constrained for simple templates, such as local, they turn computationally challenging for models with non-trivial shapes and strong scale-dependence.
This theoretical bottleneck may remain a challenge even with improvement in the observational capabilities of SKAO.

Beyond the early Universe physics, the astrophysical component of GWB serves as an independent tracer of the matter distribution.
We have discussed how cross-correlating GWB anisotropy maps with electromagnetic tracers of  LSS ($C_\ell^{\rm gal, GW}$) effectively suppresses the astrophysical shot noise of the discrete SMBHB sources~\cite{semenzato2024GWBLSS,Sah_2024,Cusin:2025xle}. 
The technique establishes GWB as a novel probe of the cosmic evolution and environmental properties of the SMBHB population.
Current PTA is fundamentally restricted by a limited number of monitored pulsars, highly non-uniform sky coverage and proper noise modeling, which limit both the effective angular resolution and accuracy of the recovered angular scales. 
The SKAO has the potential to dramatically expand PTA's capabilities, monitoring hundreds to potentially over a thousand high-precision millisecond pulsars \pcite{Keane:2014vja,Xin_2021}. 
This expanded baseline will provide near-isotropic sky coverage and enable the high-resolution GWB maps strictly required for statistically significant cross-correlations.

Even in the SKAO era, progress will depend not only on improved sensitivity, but also on the ability to identify or subtract resolvable continuous-wave sources, control the residual shot-noise contribution, and realistic modelling of the astrophysics of SMBHB population. 
The interpretation of GWB cross-correlations will further require accurate knowledge of tracer redshift distributions and biases, together with control of selection effects, magnification, and photometric-redshift systematics. For this reason, the most robust early applications will likely focus on well-characterized tracers and on angular scales where both PTA and LSS systematics are under the best control.

Weighing the possibilities and the associated caveats, SKAO is definitely transformative, representing a significant step toward using GWB to explore properties of primordial non-Gaussianities, large-scale structure, and the cosmic evolution of SMBHB population.

\section*{Acknowledgements}
DB, NB, GP \& SM acknowledge support from the COSMOS network (www.cosmosnet.it) through ASI (Italian Space Agency) Grants 2016-24-H.0, 2016-24-H.1-2018 and 2020-9-HH.0.
HVR \& NB acknowledge support by the MUR PRIN2022 Project ``BROWSEPOL: Beyond standaRd mOdel With coSmic microwavE background POLarization''-2022EJNZ53 financed by the European Union - Next Generation EU.
The work of MP is supported by the European Union under the Next Generation EU programme. MP gratefully acknowledges financial support from Fondazione Ing. Aldo Gini, from the INFN initiatives \textit{Amplitudes} and \textit{InDark}, and the hospitality and support of the Albert Einstein Institute. GP aknowledges partial financial support from Fondazione Angelo Della Riccia and Fondazione Ing. Aldo Gini. GP thanks the Leibniz Universität in Hannover for the kind hospitality for a period during which this chapter was written.

\bibliographystyle{abbrvnat-maxbibnames4}
\setlength{\bibsep}{0.0pt} 
\bibliography{chapter}

\end{document}